# Light Curves and Colors of the Ejecta from Dimorphos after the DART Impact

Ariel Graykowski[1], Ryan A. Lambert[1], Franck Marchis[1,2], Dorian Cazeneuve[1], Paul A. Dalba[1,3,4], Thomas M. Esposito[1,2,5], Daniel O'Conner Peluso[1,6], Lauren A. Sgro[1,2], Guillaume Blaclard[2], Antonin Borot[2], Arnaud Malvache[2], Laurent Marfisi[2], Tyler M. Powell[7], Patrice Huet[8], Matthieu Limagne[9], Bruno Payet[10], Colin Clarke[11,12,13] , Susan Murabana[12,14], Daniel Chu Owen[12,14], Ronald Wasilwa[12,14], Keiichi Fukui[15], Tateki Goto[16], Bruno Guillet[17], Patrick Huth[18,19], Satoshi Ishiyama[20], Ryuichi Kukita[21], Mike Mitchell[22], Michael Primm[23], Justus Randolph[24], Darren A. Rivett[25], Matthew Ryno[26], Masao Shimizu[27], Jean-Pierre Toullec[28], Stefan Will[29], Wai-Chun Yue[30], Michael Camilleri[31], Kathy Graykowski[32], Ron Janetzke[33], Des Janke[34], Scott Kardel[35,36], Margaret Loose[37], John W. Pickering[38,39,40], Barton A. Smith[41], Ian M. Transom[42,43]

Corresponding Email: agraykowski@seti.org

[1]SETI Institute, Carl Sagan Center, 339 Bernardo Ave Suite 200, Mountain View, 94043, CA, USA

[2]Unistellar, 5 allée Marcel Leclerc, Bâtiment B., Marseille, 13008, France

[3]Department of Astronomy and Astrophysics, University of California, Santa Cruz, 95064, CA, USA

[4]Heising-Simons 51 Pegasi b Postdoctoral Fellow

[5]Astronomy Department, University of California, Berkeley, 94720, CA, USA

[6]Centre for Astrophysics, University of Southern Queensland, Toowoomba, QLD, Australia

[7]Department of Earth, Planetary, and Space Sciences, University of California, Los Angeles, 90095, CA, USA

[8]Unistellar Citizen Scientist, Le Tampon, France

[9]Unistellar Citizen Scientist, Saint-Paul, Réunion

[10]Unistellar Citizen Scientist, La Rivière, Réunion

[11]Armagh Observatory and Planetarium, College Hill, Armagh BT61 9DB, United Kingdom

[12]The Travelling Telescope, Nairobi Planetarium, Kenya

[13]Unistellar Citizen Scientist, College Hill, United Kingdom

[14]Unistellar Citizen Scientist, Nairobi, Kenya

[15]Unistellar Citizen Scientist, Tsuchiura, Japan

[16]Unistellar Citizen Scientist, Osaka, Japan

[17]Unistellar Citizen Scientist, Caen, France

[18]Community College of Allegheny County, 800 Allegheny Ave #1804, Pittsburgh, 15233, PA, USA

[19]Unistellar Citizen Scientist, Schenley, PA, USA

[20]Unistellar Citizen Scientist, Chigasaki, Japan

[21]Unistellar Citizen Scientist, Kajiki Aira, Japan

[22]Unistellar Citizen Scientist, Oklahoma City, OK, USA

[23]Unistellar Citizen Scientist, Austin, TX, USA

[24]Unistellar Citizen Scientist, Athens, GA, USA

[25]Unistellar Citizen Scientist, Lake Macquarie, Australia

[26]Unistellar Citizen Scientist, Milwaukee, WI, USA

[27]Unistellar Citizen Scientist, Shimoishii, Japan

[28]Unistellar Citizen Scientist, Saint-Gilles, Réunion

[29]Unistellar Citizen Scientist, Raleigh, NC, USA

[30]Unistellar Citizen Scientist, Hong Kong

[31]Unistellar Citizen Scientist, Auckland, New Zealand

[32]Unistellar Citizen Scientist, San Francisco, CA, USA

[33]Unistellar Citizen Scientist, San Antonio, TX, USA

[34]Unistellar Citizen Scientist, Queensland, Australia

[35]Palomar Community College, San Marcos, 92069, CA, USA

[36]Unistellar Citizen Scientist, San Marcos, CA, USA

[37]Unistellar Citizen Scientist, San Diego, CA, USA

[38]Department of Medicine, University of Otago, Christchurch, New Zealand

[39]Emergency Care Foundation, Christchurch Hospital, Christchurch, New Zealand

[40]Unistellar Citizen Scientist, Christchurch, New Zealand

[41]Unistellar Citizen Scientist, Campbell, CA, USA

[42]Hamilton Astronomical Society Observatory, 183 Brymer Road, Rotokauri, Hamilton, 3200, New Zealand

[43]Unistellar Citizen Scientist, Cambridge, New Zealand

**On 26 September 2022 the Double Asteroid Redirection Test (DART) spacecraft impacted Dimorphos, a satellite of the asteroid 65803 Didymos[1]. Because it is a binary system, it is possible to determine how much the orbit of the satellite changed, as part of a test of what is necessary to deflect an asteroid that might threaten Earth with an impact. In nominal cases, pre-impact predictions of the orbital period reduction ranged from ~ 8.8 – 17.2 minutes[2,3]. Here we report optical observations of Dimorphos before, during and after the impact, from a network of citizen science telescopes across the world. We find a maximum brightening of 2.29 ± 0.14 mag upon impact. Didymos fades back to its pre-impact brightness over the course of 23.7 ± 0.7 days. We estimate lower limits on the mass contained in the ejecta, which was 0.3 – 0.5 % Dimorphos' mass depending on the dust size. We also observe a reddening of the ejecta upon impact.**

Four Unistellar eVscopes captured observations of Didymos during the DART impact into Dimorphos on the night of 26 September 2022. Of all telescopes that observed the DART impact, from the ground and space, the eVscopes were among the smallest with apertures of 112 mm. Three eVscopes were located on Réunion Island and one in Nairobi, Kenya. Figure 1 shows eVscope images taken before and after the impact as well as the ejecta produced by the impact. We observe a fast-moving ejecta plume moving eastward on the plane of the sky and spreading over a timescale of minutes, as well as slower moving ejecta that morphologically appear to have formed a coma.

Of the four eVscope data sets that included the moment of impact, three were suitable for photometric analysis because the observers recorded dark frames for image calibration. Thus, we conducted aperture photometry on these three data sets through a circular aperture radius of 13.6" or 750 km at the distance of Didymos The resulting apparent magnitudes measured over time are displayed in Figure 2.

Using data from the eVscopes located in L'Étang-Salé and Saint-Paul, Réunion, we calculate an impact time of UTC 23:15:02 ± 4 s on 2022 Sep 26, which agrees with the reported Earth-observed impact time of 23:15:02.183 UTC [1], itself coming 38 seconds after the actual time of impact on the spacecraft due to light-travel time (private communication with Julie Bellerose). Before the impact, we use the observations taken with the eVscope located in L'Étang-Salé, Réunion to measure an apparent visual magnitude, $m_v$ = 14.48 ± 0.11 and a minimum magnitude (maximum brightness) $m_v$ = 12.18 ± 0.03. As the fast-moving ejecta moved out of the photometric aperture, the magnitude increased to $m_v$ = 12.96 ± 0.04. At a geocentric distance of 0.076 au, a heliocentric distance of 1.05 au, and a phase angle of 53.28° at the time of impact, these correspond to absolute visual magnitudes of $H_v$ = 18.12 ± 0.11, 15.83 ± 0.03, and 16.61 ± 0.04 respectively.

We estimate the effective mass in the fast-moving ejecta by calculating the change in effective cross sections corresponding to magnitudes $H_v$ = 15.83 ± 0.03 and 16.61 ± 0.04, and assuming an average density and albedo for Dimorphos. This is further detailed in the "Mass of the Ejecta" section of the Methods. We must also assume an average particle radius, so we consider several scenarios.

To begin with, the particles must be small as evidenced by the long tail that developed in the anti-solar direction several days after impact. We approximate an antisolar tail length of ~7 × $10^3$ km on the plane of the sky ~ 113.7 hours after impact as shown in Figure 1. Measurements from the 4.1-meter Southern Astrophysical Research (SOAR) Telescope, at NSF NOIRLab's Cerro Tololo Inter-American Observatory in Chile revealed the tail length to be >$10^4$ km on the plane of the sky two days after impact[4]. Our measurement is shorter due to the smaller collecting area of the eVscope and its lesser sensitivity to low surface brightness. We measure a 3σ limiting magnitude of 17.01 ± 0.03 in this image stack consisting of 1,205 four-second exposures.

We then refer to active asteroid (596) Scheila, whose activity was likely caused by an impact and whose average ejected dust size is predicted to be small based on the observed effect of solar radiation sweeping[5]. From this, average dust radii spanning $\overline{a}$ ~ 0.1 - 1.0 $\mu$m were estimated. We point out that larger dust sizes were also estimated for Scheila such as $\overline{a}$ ~ 100 $\mu$m based on a modeled particle range of $a$ ~ 1 to ~ $10^4$ $\mu$m along a power-law distribution with an index $q$ = 3.5[6,7].

We first examine an average particle radius $\overline{a}$ ~ 1 $\mu$m, as particles much smaller than this become less efficient at scattering visible light, and particles much larger than this are expected to persist longer in the photometric aperture. This particle radius then results in a mass of $m_{\mathrm{fe}}$ ~ (7.0 ± 1.2) × $10^3$ kg contained in the fast-moving ejecta plume respectively. We measured the speed of this fast-moving ejecta on the plane of the sky through the photometry aperture and find this fast-moving ejecta has a speed of $v_{fe}$ ~ 970 ± 50 m $s^{-1}$ . The resulting mass then corresponds to a kinetic energy of $KE_{fe}$ ~ (3.3 ± 0.6) x $10^9$ J. The relative kinetic energy of the DART spacecraft at the time of impact is $KE_D \sim (1.094 \pm 0.001) \times 10^{10}$ J [1]. This implies that the observed fast-moving ejecta plume carried away ~30 ± 6% of the kinetic energy delivered by the DART spacecraft. This is comparable to impact simulations, which have shown that kinetic energy can be transferred from the impactor to the ejecta on the order of tens of percent[8, 9, 10]. Average particle sizes an order of magnitude larger than this would exceed the kinetic energy introduced by the DART spacecraft, so we do not further consider larger particles sizes as making up a significant portion of the observed fast-moving ejecta plume. The lower end of the dust size estimations based on solar radiation pressure, $\overline{a} \sim 0.1\ \mu$m, results in an estimated mass of $m_{\mathrm{fe}} \sim (7.0 \pm 1.2) \times 10^2$ kg, corresponding to ~3.0 ± 0.6% of DART's kinetic energy. These values likely represent underestimates on ejecta mass, given the assumed average dust radii, as only the particles with velocities high enough to escape Dimorphos ($v > v_e \sim 0.087 \pm 0.01$ m s$^{-1}$) will contribute to the measurable increase in cross section[1, 11, 12].

To estimate the mass of ejecta contained within the coma, we measure Didymos' fading rate. The absolute magnitudes of Didymos after impact are plotted in Figure 3. An error-weighted linear fit indicates that it took ~23.7 ± 0.7 days for the dust to move out of the photometric aperture with velocity $v_{ce} \sim 0.37 \pm 0.01$ m s$^{-1}$. We use an approximated impact relation (Equation 6 in Methods)[13] to relate this with the mass, density, and speed of the DART impactor, and estimate the mass contained in the coma that resulted from the impact to be $m_{\mathrm{ce}} \sim (1.3 \pm 0.1) \times 10^7$ kg. Considering the change in effective cross section before impact and after the fast-moving ejecta dissipated, Equation 3 in the Methods section results in and average dust radius $\overline{a} \sim 1.7 \pm 0.3$ mm.

We also estimate the particle radius in the observed coma by relating solar radiation pressure to the turn-around distance of the particles in the coma along the comet-Sun line, and the bulk velocity of the particles, $v_{ce}$ (detailed in the "Mass of the Ejecta" section of the Methods). It was found that particles reached distances d ~ 150 - 250 km in the solar direction before turning around[14]. We, then find average dust radii $\overline{a}$ ~ 2.8 ± 0.3 – 3.8 ± 0.4 mm in the coma, corresponding to a mass range of $m_{ce}$ ~ (1.3 – 2.2) × $10^7$ kg, in good agreement with our estimate above.

In all scenarios, we find that while the impact led to a significant increase in apparent magnitude, the overall mass loss in the observed fast-moving ejecta plum or slower moving ejecta in the coma created by DART's impact into Dimorphos was not totally disruptive.

We also measure the color before, during, and after DART's impact into Dimorphos as seen in Figure 4 to show a significant reddening at the time of impact. The colors appear to return to their original color as the fast-moving ejecta plume dissipates. This initial reddening was also seen on 9P/Tempel 1 after the impact of NASA's *Deep Impact* spacecraft and was determined to be caused by different sized particles having a range of velocities, causing the particle size distribution and the ejecta optical depth to change over time[15].

If the reddened color is interpreted as a proxy for material composition, we must consider this in the context of the colors of active and inactive small bodies in the solar system. Typically, active bodies appear bluer in color on average than their inactive counterparts, such as short-period comets versus Kuiper belt objects[16]. Some of these redder observed surface colors may be due to irradiation of organics[17], which can efficiently mantle the surface of bodies like the Kuiper belt objects[18,19]. Additionally, in general, the highest velocity ejecta from an impact is sourced from the material closest to the site of impact[20]. The fast-moving ejecta plume may be more representative of the surficial material of Dimorphos than the material in the slower-moving coma. While much of the fine-grain surface regolith on Dimorphos has likely been depleted through processes like electrostatic removal[21], it is possible that the remaining small particles have experienced some irradiation. Spectra indicate the presence heterogeneity of surface materials on Didymos, with various concentrations of likely hypersthene with a grain radius $a$ <25 $\mu$m, and olivine with a grain radius $a$ <45 $\mu$m [22]. Lab simulations of space weathering showed that irradiation can cause significant reddening of olivine's reflectance spectrum[23]. We emphasize that these posed interpretations are a few of possibly many.

## Figure Legends

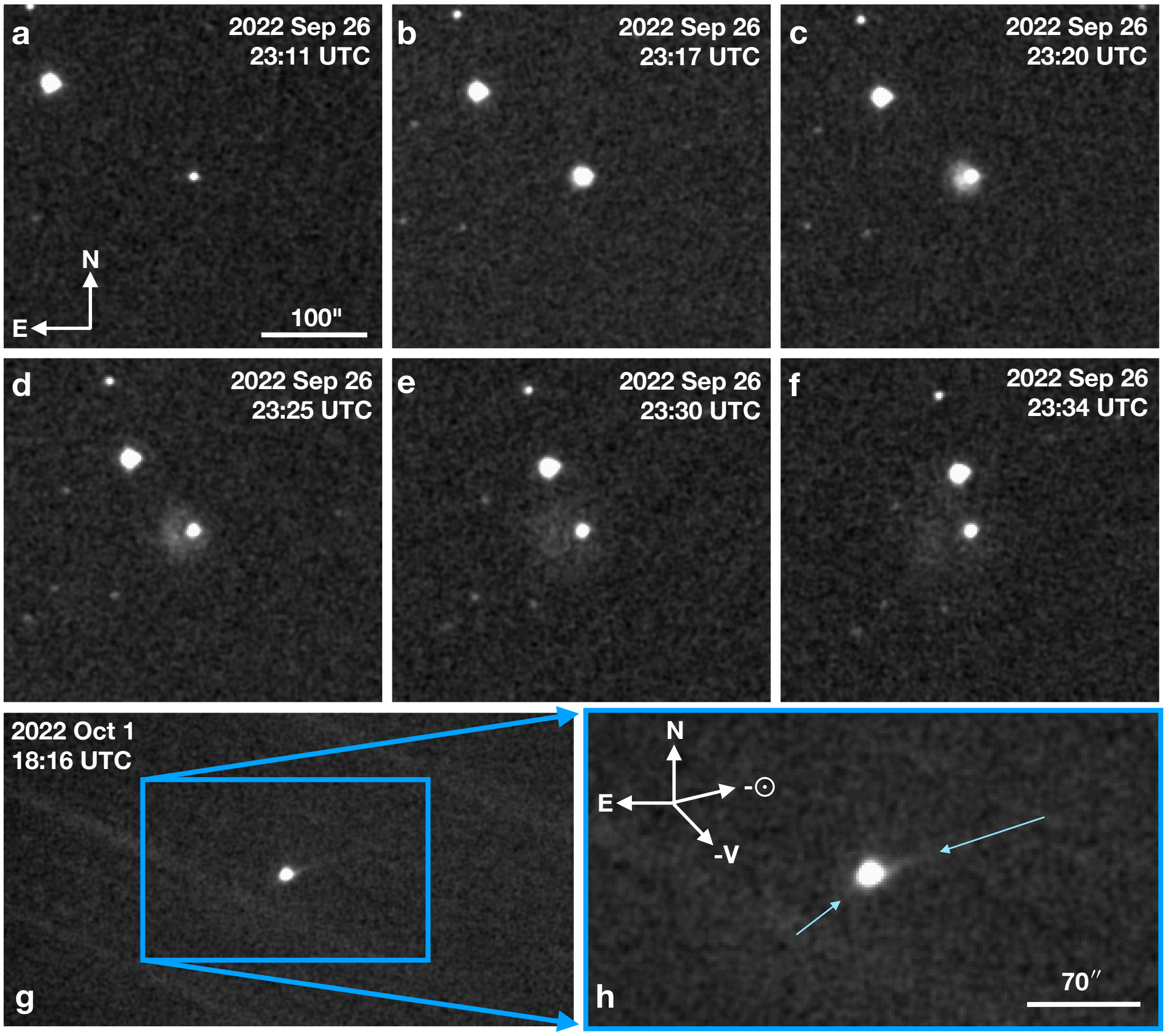

**Figure 1**. **eVscope Observations of the Impact, Ejecta, and Tail**. Panels a-f show the initial ejecta plume from DART's impact into Dimorphos as observed from L'Étang-Salé, Réunion. Panel a shows Didymos before the DART impact. Panels b-f show Didymos after the DART impact. The compass and image scale in panel a applies to panels b-f as well. The fast-moving ejecta plume moves eastward on the plane of the sky and dissipates over time, from panels a to f. Panels g and h show two tails (~solar and ~anti-solar directions) developed from the ejecta produced by the DART spacecraft ~ 113.7 hours after impact into Dimorphos. The image in panel h is a zoomed-in version of the image in panel g. This image is a median stack of 1,205 four-second exposures as observed from Nagahama, Japan. The two light blue arrows mark the two tails visible to an eVscope as visual aids.

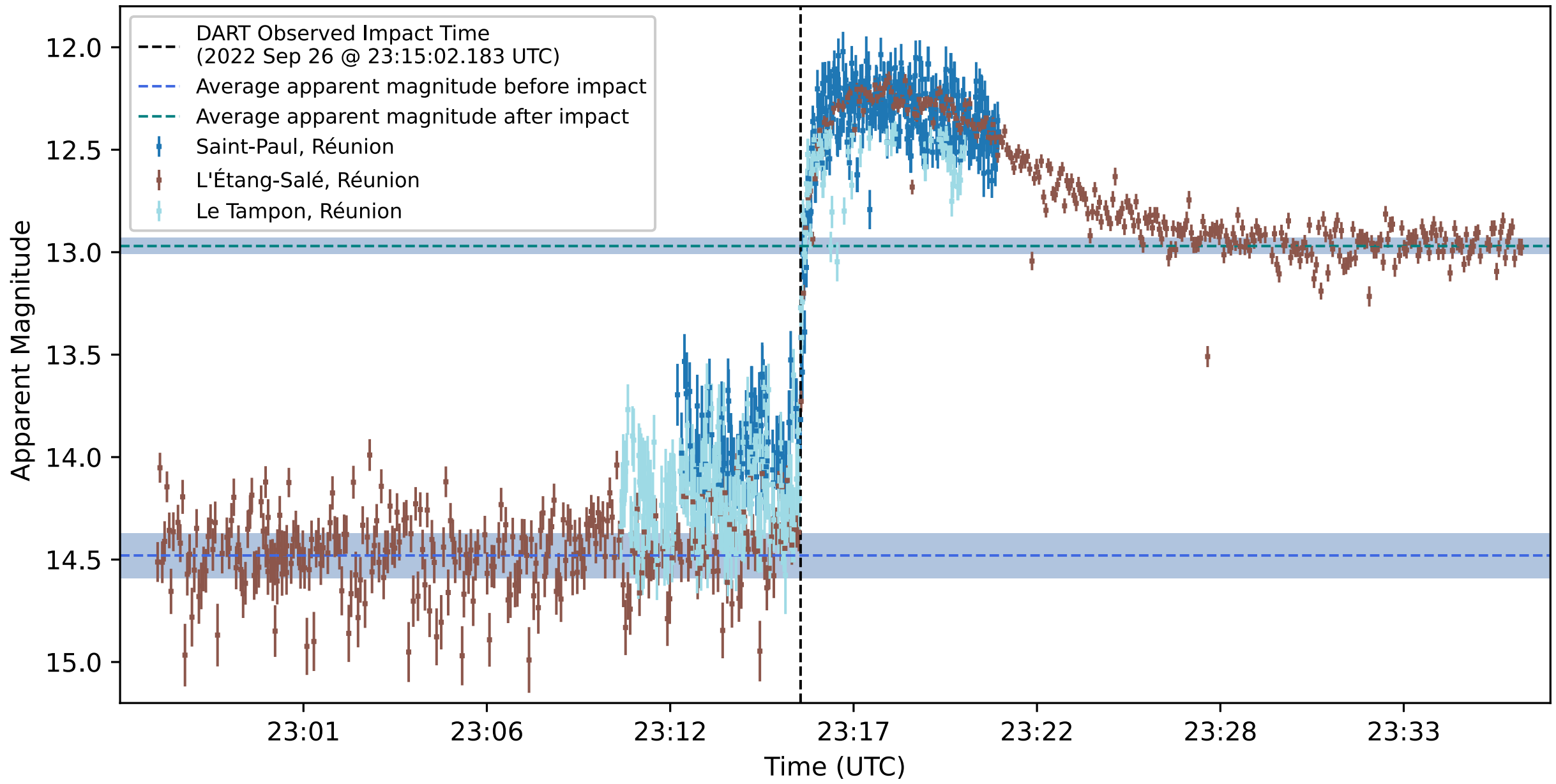

**Figure 2. Apparent Magnitude of Didymos Before, During, and After Impact**. The light curve of the Didymos binary system on 26 September 2022 during the DART spacecraft's impact into Dimorphos as observed by three citizen astronomers located on Réunion Island using eVscopes. The dotted lines are the measured apparent magnitude before impact (bottom) and after the fast-moving ejecta dissipated after the impact (top). The shaded regions represent the standard deviation on the value of the dotted lines, and error bars represent sky background noise.

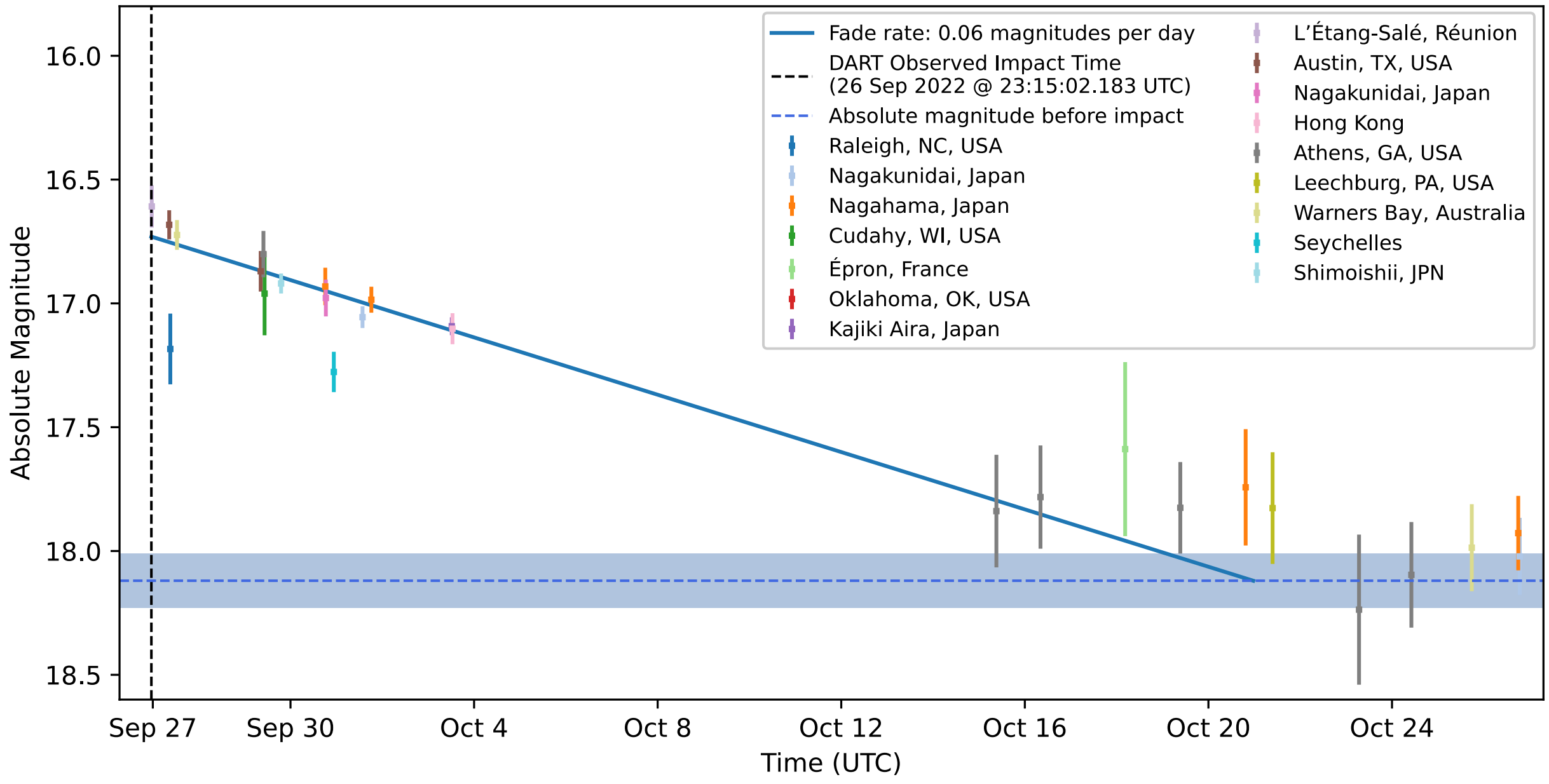

**Figure 3**. **Fading of Didymos After Impact.** The absolute magnitude of the Didymos system faded over time after the brightening due to DART's impact into Dimorphos. The solid blue line is a weighted, linear fit to magnitudes measured from just after impact on 2022 Sep 26 through 2022 Oct 25, after which measurements were consistent with the resting absolute magnitude. Some measurements between Oct 15 and 25 overlap the resting magnitude at the ~1σ level but others remain above it. Therefore, we consider the fading time may range between ~18 and ~28 days after impact, with our best-fit model providing a fading time of 23.7 ± 0.7 days. The value of the resting absolute magnitude is calculated from the pre-impact average apparent magnitude plotted in Figure 2. The error bars and shaded region represent the 1σ measurement uncertainties. Prior to 2022 Oct 4, there are two outlying observations that resulted in measurements that were too faint due to poor weather conditions, and we therefore do not include these points in the fitted line.

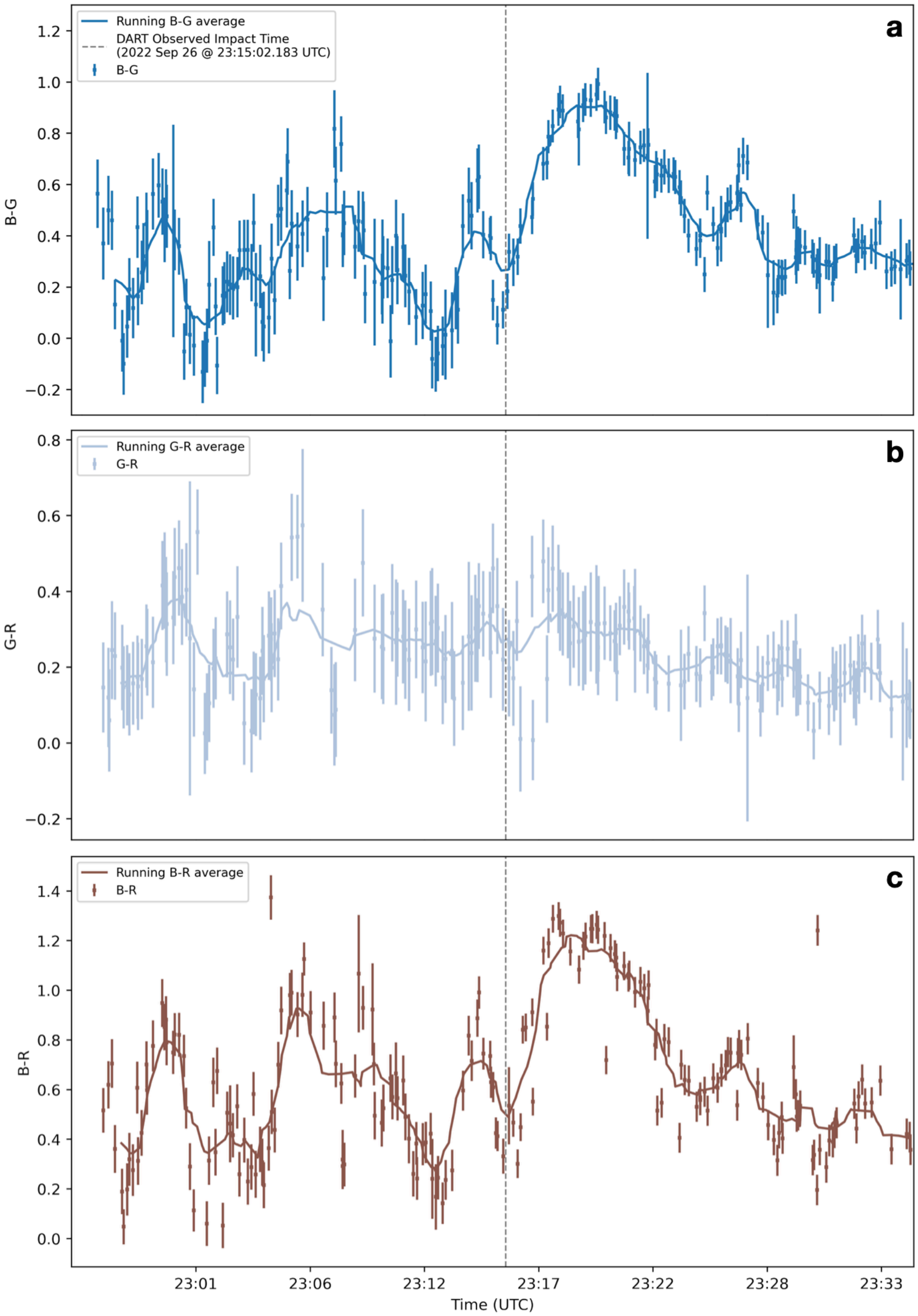

**Figure 4**. **Colors of Didymos Before, During, and After Impact**. Measured B-G (a), G-R (b), and B-R (c) colors of Didymos over time as observed from L'Étang-Salé, Réunion. Error bars represent the noise introduced by the sky background magnitude. It is apparent in panels a and c that the Didymos binary system became redder directly after the DART spacecraft impacted Dimorphos and that when the fast ejecta dissipated, the colors returned to their original colors. The G-R colors in panel b show no significant change in color during the time of impact.

# Methods

## Observations

The measurements in this work are derived from 23,240 images taken by 17 different eVscopes in the Unistellar network of eVscopes. At the time of this study, the Unistellar fleet of eVscopes contained three variations: the eVscope 1, eVscope eQuinox, and the eVscope 2. All eVscopes have 112 mm diameter primary mirrors. The eVscope 1 and eQuinox both utilize a Sony IMX224LQR sensor and have a 37'× 28' field of view and pixel scale of 1.72"/pix, while the eVscope 2 utilizes a Sony IMX347LQR sensor and has a 45'× 34' field of view and pixel scale of 1.33"/pix. The detectors have RGB photosensors arranged as Bayer filter matrices. Exposure times of individual images varied between observers from 1.0 to 4.0 s but remained constant throughout each individual data set. Images were taken consecutively over ranges of 10 to 80 minutes. In this work, we report observations taken from 2022 Sep 26 through 2022 Oct 31 UT. Over this time range, the observing parameters of Didymos ranged from geocentric distances of 0.076 to 0.109 au, heliocentric distances of 1.046 to 1.018 au and phase angles of 53.2° to 73.8°. Figure 1 shows two of the tails that formed after impact as seen on 2022 Oct 1, when the position angle was 60.4° in the projected anti-solar direction and 27.6° in the projected anti-velocity direction. In each data set, the limiting magnitude is between 15 and 16 magnitudes, except for the stacked image that shows the tails in Figure 1, which consists of 1,250 four-second images. For this data set, we measure a 3σ limiting magnitude of 17.01 ± 0.03. We further summarize the observation circumstances in Extended Data Table 1.

**Aperture Photometry**

To measure the magnitude of Didymos in the images, we first reduce the data submitted by Unistellar citizen astronomers. Unistellar Citizen astronomers submit science images with their chosen parameters (as listed above) as well as dark frames for calibration which are recorded immediately after the science observations. The dark frames are median combined and subtracted from the science images. The Unistellar Network is currently not prompted to take flat frames to further calibrate the data. In general, we find the errors on photometric measurements are dominated by photon noise and sky rather than the lack of a flat-field correction. We also aim to keep Didymos in the middle of the detector when conducting observations to further mitigate the lack of a flat frame calibration. Before conducting aperture photometry, the science images are aligned and median-combined in time; however, the observations conducted during the DART impact displayed in Figure 2 were measured from images that were not stacked because of the speed of brightening upon impact. The photometry measurements on the observations conducted after the impact were conducted on median-combined images that vary from 17 to 30 images. The measurements are then averaged and plotted at the midpoint in time of the conducted observation. We choose an aperture radius of 13.6" corresponding to 750 km at the distance of Didymos) through which we measure the flux. The PSF of Didymos after the impact depends on the dataset. Table 1 in Extended Data lists the average full width at half maximum (FWHM) of the gaussian-fitted PSFs of the stars in each dataset. The most egregious cases were due to out of focus eVscopes. We find the FWHM of the PSF of Didymos is ~9.0 ± 1.0% larger than that of the stellar PSFs. Our chosen aperture is large enough to measure all the flux from the Didymos system throughout all the submitted datasets. We subtract the flux of the sky background measured through an annulus centered on the target that ranges from 22.21" to 30.75" in radius, corresponding to ~1218 km to ~1686 km at the distance of Didymos. We use stars in the field with known Gaia magnitudes to convert our instrumental flux values to apparent magnitudes, $m_V$. These magnitudes are plotted in Figure 2. When measuring the magnitude before impact and after the fast-moving ejecta dissipated, we use observations from the eVscope in L'Étang-Salé, Réunion on 2022 Sep 26. This eVscope was able to reach the greatest sensitivities as it was contained in a protective dome that blocked wind and stray light.

To obtain the colors of the Didymos binary system, the color channels of the detectors must be isolated. In a Bayer filter matrix of blue (B), green (G), and red (R) photosensors, the first 2 × 2 section of pixels is arranged as [R, G; G, B]. This pattern repeats to fill the size of the detector. We isolate each color channel based on this known pattern of the photosensors and conduct aperture photometry as described above. We calibrate the instrumental fluxes using calibrator stars with known Gaia magnitudes in the background of each image. We measure the aperture fluxes of those calibrator stars in each color channel and, combined with their known Gaia magnitudes, use them to convert the Didymos aperture fluxes in those channels to B, G, and R magnitudes[24]. We plot the resulting B-G, G-R, and B-R colors, the B-G and B-R colors in Figure 4.

**Mass of the Ejecta**

Apparent magnitudes $m_V$ are converted to absolute magnitudes, $H_V$, which represents the magnitude of the object with heliocentric and geocentric distance ($r_H$ and $\Delta$ respectively) of 1 au and at phase angle, $\alpha = 0°$. The correction is

$$H_V = m_V - 5\log_{10}(r_H\Delta) - \beta\alpha \tag{1}$$

where $\beta\alpha$ is the phase function that represents the dependence on sunlight scattering by the dust particle at an angle $\alpha$ in degrees. We assume a linear phase function with phase coefficient, $\beta$ = 0.035, a typical value for S-type asteroids like Didymos[25, 26, 27, 28, 29]. With this, our measured absolute magnitude before the impact of $H_V$ = 18.12 ± 0.11 agrees with past measurements of $H_V$ = 18.16 ± 0.04 of the binary system[30]. Absolute magnitudes are plotted in Figure 3. Next, we estimate the effective cross sections with

$$C_e = \frac{\pi(2.25\times10^{16})}{p_v} 10^{-0.4}\left[H_V - V_{\odot}\right] \tag{2}$$

where $p_v$ ~ 0.15 ± 0.02 is the albedo of the Didymos system measured by the DRACO camera[1], $H_V$ is the absolute magnitude we measure for the Didymos system, and $V_{\odot}$ ~ -26.77 is the apparent magnitude of the Sun. For Didymos, we find an effective scattering cross section $C_e$ ~ 0.53 ± 0.06 km$^2$ before impact, $C_e$ ~ 4.35 ± 0.13 km$^2$ at Didymos' peak in brightness just after impact, and $C_e$ ~ 2.17 ± 0.09 km$^2$ after the fast-moving ejecta moved out of the photometric aperture. A cross section $C_e$ ~ 0.53 ± 0.06 km$^2$ implies an effective radius $r_e$ ~ 411 ± 22 m, which is consistent with previous radar measurements of Didymos finding a volume equivalent radius of $r_e$ ~ 390 ± 15 m [11]. The mass is related to the effective cross section by

$$M_e = \frac{4}{3}\rho\bar{a}C_e \tag{3}$$

where $\rho$ ~ 2400 ± 250 kg $m^{-3}$ is the bulk density of the Didymos system[1], and we adopt $\bar{a} = \sqrt{a_{min}a_{max}}$ for the mean dust radius amongst particles having a size range $a_{min} \leq a \leq a_{max}$. We examine particles with mean radii $\bar{a}$ ~ 0.1 - 1 $\mu$m in the fast-moving ejecta as was found for the ejecta in impacted asteroid Scheila[5]. The change in effective cross section measured at peak brightness and at the leveled brightness after the fast-moving ejecta dissipated then allows us to measure the change in mass, or the mass contained in the fast-moving ejecta plume with Equation 3. We do the same for the mass lost in the coma of slower moving particles by examining the change in cross section from before impact and after impact, when the fast-moving ejecta dissipated from the photometric aperture. However, instead of assuming a dust size within the coma, we estimate the dust size in two ways. First, we estimate the mass in the coma based on the fading time of Didymos after impact and the speed of the particles through a process explained in the following section. Equation 3 then gives an estimate on the average dust radius. We also estimate the dust radius by connecting the distance a particle can reach in the Solar direction before being turned around by Solar radiation pressure to its initial speed using

$$B = \frac{u^2 {r_H}^2}{2GM_{\odot}\ell} \tag{4}$$

where $B$ is a dimensionless radiation pressure factor, $G = 6.67 \times 10^{-11}$ m$^3$ kg$^{-1}$ s$^{-2}$ is the gravitation constant, $M$ is the mass of the Sun, $r_H \sim 1.04$ AU is the heliocentric distance of Dimorphos on 2022 Oct 1, when the turn-around distance was measured to reach $\ell$~150-200 km [14], and $u$ is the initial velocity of the particles which we measure as $u \sim v_{ce} \sim 0.37 \pm 0.01$ m s$^{-1}$ from the fading rate of Didymos. $B$ is the ratio of acceleration due to solar radiation pressure to the acceleration due to solar gravity expressed as

$$B = \frac{KQ_{pr}}{\rho a} \tag{5}$$

where $K = 5.7\times 10^{-4}$ kg m$^{-2}$ is a constant, $Q_{pr}$ is the radiation pressure coefficient often assume to be 1, $\rho$ is the particle density, and we examine the average dust $a = \overline{a}$. Again, we examine dust densities equivalent to the bulk density of the Didymos system[1]. Assuming $\rho \sim 2400 \pm 250$ $kg\ m^{-3}$ gives $B \sim (0.24 \pm 0.02) / \overline{a}$ microns. We point out that using a particle $\rho \sim 3480 \pm 80$ $kg$ $m^{-3}$, which is the average density of LL ordinary chondrite material[31] as associated with S-type asteroid Didymos[32], also result sin mm size particles corresponding to masses within the range we estimate when assuming $\rho \sim 2400 \pm 250$ kg $m^{-3}$.

**Speed and Energy of the Ejected Dust**

To estimate the energy carried by the mass in that fast-moving ejecta plume and slower moving coma, we obtain their speeds. The fast-moving ejecta can be measured visually, following the plume on the detector over time. Additionally, we measure the crossing time, $t_c$, of the particles in the photometric aperture. This is the time between the moment of impact and the moment the magnitude settled to $m_v = 12.96 \pm 0.04$. We determine the peak time and dissipation time of the fast-moving ejecta by analyzing the magnitudes binned into rolling bins of 5 images to determine peak time and 15 images for the settling time. We then choose times when the residuals were within the respective measured errors on the peak magnitude and magnitude after fast-moving ejecta dissipation. We find a crossing time $t_c \sim 775 \pm 40$ s over a photometric aperture radius of 13.67", which is equivalent to ~750 km at the distance of Didymos. Therefore, we obtain a speed of $v \sim 970 \pm 50$ m s$^{-1}$.

In the slower moving ejecta that makes up the coma, we estimate the particle speed from the fading time of Didymos after impact. This is the time between the moment of impact and the moment the magnitude increased back to Didymos' original absolute magnitude, $H_V$ = 18.12 ± 0.11. Then, assuming equal projectile and target densities, we make a simple estimation of the mass of the ejecta $m_e$ by the impact relation

$$m_e = A * M_P \left(\frac{u}{U_P}\right)^s \tag{6}$$

where $U_P$ ~ 6144.9 ± 0.3 is the impact speed of the DART spacecraft[1], $M_P$ ~579.4 ± 0.7 is the mass of the DART spacecraft on impact[1], $u$ ~ $v_{ce}$ ~ 0.37 ± 0.01 m $s^{-1}$ is the bulk velocity of the particles in the ejecta, and for consistency with past works A = 0.01 is a constant and s is an index that that we approximate to $s$ = -1.5 but depends on the material[7, 13]. The density of Dimorphos is assumed to be the same as the bulk density of the Didymos system in the calculations, however we emphasize that the density of Dimorphos alone is not measured. The density of the main body of the DART spacecraft (without the solar panels) was ~ 270 kg $m^{-3}$ at impact time[33]. For the sake of this simple approximation, we take the densities of the spacecraft and Dimorphos to be similar enough, because we expect lower densities with decreasing diameter for S-type asteroids[34] and that the Spacecraft was stopped by Dimorphos rather than flying straight through it. We also emphasize that this approximation comes with the caveat that impact physics on small asteroids are still not well-understood, so this impact relation serves as a rough estimate[34].

With the estimated masses, $m$ and speeds, $v$ of the ejecta, we can estimate the kinetic energy, $E_K$ carried by the initial fast-moving ejecta as well as the coma by $KE = (1/2)\, mv^2$. As a tool to choose appropriate dust size test-cases, we compared the estimated kinetic energies to the kinetic energy introduced by the DART spacecraft where mass on impact is $M_P$ ~579.4 ± 0.7 kg and velocity on impact is $v = U_P \sim 6144.9 \pm 0.3$ m s$^{-1}$. We also compare this to the orbital energy, $E_O$ of Dimorphos around Didymos before and after the orbital period change of -33.0 $\pm$ 1.0 minutes[35] using

$$E_O = \frac{-GM_{Did}m_{Dim}}{2r} \tag{7}$$

where $M_{Did} \sim 5.6 \pm 0.5 \times 10^{11}$ kg, $m_{Dim} \sim 4.3 \times 10^{9}$ kg, and r is the semi-major axis of the orbit of Dimorphos around Didymos[1,36]. Before the impact, the semi-major axis was measured to be r = 1.206 $\pm$ 0.035 km [1]. We can estimate the original orbital energy of Dimorphos to be $E_O \sim -(6.6 \pm 0.6) \times 10^{7}$ J. With the orbital period decrease, Kepler's third law gives a new semimajor axis of r ~ 1.2 $\pm$ 0.1 km resulting in a change of orbital energy of $\Delta E_O \sim (2.1 \pm 0.6) \times 10^{6}$ J. These simple approximations are sufficient for the scope of this work, which aims to use these estimates as a means of providing a check on the reasonability of the estimated masses of the ejecta.

## Data Availability

The Unistellar network of citizen astronomers have the option to upload their FITS images to an AWS server rented by the Unistellar corporation. This data is then available upon request. The resulting photometry used in the data analyses are made available on the corresponding author's public GitHub repository in the form of CSV files corresponding to the figures and extended data table. The repository also contains the FITS images used in Figure 1. The data is located in the corresponding author's public GitHub repository (https://github.com/Ariel-Graykowski/DART_Unistellar)[37].

## Code Availability

The SETI/Unistellar pipeline used to dark-subtract and stack the astronomical images and conduct aperture photometry is currently located on Unistellar's private GitHub. These codes are available upon request. Python codes used to create the figures and conduct the main data analysis are located in the corresponding author's public GitHub repository (https://github.com/Ariel-Graykowski/DART_Unistellar)[37].

## Methods References

## Inclusion & Ethics

Citizen astronomers, who are listed as co-authors, voluntarily observed this event and uploaded their data to an Amazon Web Services (AWS) server rented by the Unistellar Corporation. The analyses presented in this work were conducted locally at the SETI Institute, while the observations used in this work were conducted by citizen astronomers globally. This global network of eVscopes was a necessary tool to coordinate the proper timing and location to observe the DART impact and to continuously monitor the Didymos system over time.

## Acknowledgements

This work was supported by a generous donation from the Gordon and Betty Moore Foundation. We thank Julie Bellerose for their communication on the observed timing of the DART impact. We thank the anonymous referees for reading the manuscript and providing helpful comments.

## Author Contributions

A. G., R. A. L., and F. M. initiated this work. F. M. envisioned and organized the contribution of Unistellar network for the observations of the DART impact. A. G. and R. A. L. conducted photometric measurements on the images. A. G. led the analysis of the photometric datasets and R. A. L., F. M. provided significant contribution. R. A. L., F. M., D. C., P. A. D., T. M. E., D. O'C. P., L. A. S. contributed to the interpretation, discussion, and the writing as well. F. M., A. M., L. M., and A. B. contributed to the creation of the Unistellar network. G. B. contributed to the establishment of the photometric pipeline. T. M. P. contributed to discussion of impact physics. P. H. (1), M. L., B. P., C. C., S. M., D. C. O., and R. W. contributed observations of Didymos taken before, during and after the impact of DART. M. M., K. F., T. G., B. G., P. H. (2), S. I., R. K., M. P., J. R., D. A. R., M. R., M. S., J-P. T., S. W., W-C. Y., and P. G. contributed observations of Didymos in the weeks following the impact in this work. M. C., K. G., R. J., D. J., S. K., M. L., J. W. P., B. A. S., and I. M. T. contributed observations of Didymos before the DART impact, which were used to prepare observation parameters for impact, and prepare the analyses that were used in this paper.

## Competing Interests

The authors declare no competing interests.

## Additional Information

Correspondence should be addressed to Dr. Ariel Graykowski (agraykowski@SETI.org)

# Extended Data

| Observer Initials | Observation start time (UTC) | Duration (min) | Location | Exposure time (s) | # Frames | PSF FWHM (arcsec) |
|---|---|---|---|---|---|---|
| P. H.(1) | 2022 Sep 26 23:09:49 | 10 | Le Tampon, Réunion | 1 | 596 | 7.0 ± 0.4 |
| B. P. | 2022 Sep 26 22:54:32 | 40 | L'Étang-Salé, Réunion | 4 | 624 | 6.29 ± 2.0 |
| M. L. | 2022 Sep 26 23:10:45 | 10 | Saint Paul, Réunion | 1 | 597 | 5.9 ± 0.4 |
| M. P. | 2022 Sep 27 08:10:07 | 50 | Austin, TX, USA | 4 | 759 | 18 ± 2.0 |
| S. W | 2022 Sep 27 09:04:06 | 10 | Raleigh, NC, USA | 4 | 103 | 5.9 ± 0.2 |
| D. R. | 2022 Sep 27 11:39:57 | 80 | Warners Bay, Australia | 4 | 1059 | 5.6 ± 0.2 |
| M. P. | 2022 Sep 29 08:03:06 | 40 | Austin, TX, USA | 4 | 568 | 10.26 ± 2.5 |
| J. R. | 2022 Sep 28 08:47:46 | 180 | Athens, GA, USA | 4 | 1243 | 13.6 ± 0.4 |
| M. R. | 2022 Sep 29 10:08:40 | 40 | Cudahy, WI, USA | 4 | 193 | 9.3 ± 0.4 |
| M. S. | 2022 Sep 29 18:20:08 | 90 | Shimoishii, Japan | 4 | 1279 | 8.6 ± 0.3 |
| T. G. | 2022 Sep 30 17:30:08 | 80 | Nagahama, Japan | 4 | 1231 | 3.9 ± 0.6 |
| K. F. | 2022 Sep 30 17:55:15 | 110 | Nagakunidai, Japan | 4 | 1557 | 6.0 ± 0.4 |
| J.-P. T. | 2022 Sep 30 22:36:44 | 6 | Seychelles | 4 | 95 | 4.4 ± 0.7 |
| T. G. | 2022 Oct 01 16:55:38 | 120 | Nagahama, Japan | 4 | 1217 | 6.8 ± 0.4 |
| K. F. | 2022 Oct 01 17:25:11 | 80 | Nagakunidai, Japan | 4 | 866 | 6.3 ± 0.35 |
| R. K. | 2022 Oct 03 15:58:01 | 90 | Kajiki Aira, Japan | 4 | 578 | 4.1 ± 0.6 |
| W.-C. Y. | 2022 Oct 03 16:20:05 | 220 | Hong Kong | 4 | 1345 | 3.7 ± 0.5 |
| J. R. | 2022 Oct 14 08:54:15 | 30 | Athens, GA, USA | 4 | 425 | 13.4 ± 1.5 |
| J. R. | 2022 Oct 15 07:54:53 | 30 | Athens, GA, USA | 4 | 468 | 5.7 ± 0.3 |
| B. G. | 2022 Oct 17 03:29:54 | 110 | Épron, France | 4 | 1425 | 5.2 ± 0.5 |
| J. R. | 2022 Oct 18 09:03:07 | 30 | Athens, GA, US | 4 | 176 | 5.1 ± 0.5 |
| K. F. | 2022 Oct 20 19:01:10 | 40 | Nagakunidai, Japan | 4 | 561 | 6.8 ± 1.0 |
| P. H. (2) | 2022 Oct 20 08:38:31 | 130 | Leechburg, PA, USA | 4 | 1871 | 7.3 ± 0.8 |
| J. R. | 2022 Oct 22 06:32:53 | 30 | Athens, GA, USA | 4 | 469 | 5.4 ± 0.5 |
| J. R. | 2022 Oct 23 09:52:59 | 30 | Athens, GA, USA | 4 | 469 | 4.2 ± 0.4 |
| D. A. R. | 2022 Oct 25 17:25:51 | 40 | Warners Bay, Australia | 4 | 457 | 5.9 ± 0.2 |
| T. G. | 2022 Oct 26 17:51:37 | 30 | Nagahama, Japan | 4 | 481 | 4.9 ± 0.2 |
| K. F. | 2022 Oct 26 18:28:15 | 40 | Nagakunidai, Japan | 4 | 658 | 5.9 ± 0.4 |
| M .M. | 2022 Oct 30 06:45:23 | 130 | Oklahoma City, OK, USA | 4 | 1869 | 5.7 ± 0.6 |

**Extended Data Table 1. Summary of observations.** Listed are the observations used in the above analyses from 2022 Sep 2022 to 30 Oct 2022. Observations were conducted by citizen astronomers across the World. The FWHM of the point spread function (PSF) of the stars is representative of the average "seeing" measured in each set of observations.